\documentclass{PoS}

\PoS{PoS(jhw2004)012}

\title{Modified Gravity vs Dark Matter: Relativistc theory for MOND}

\ShortTitle{Modified Gravity vs Dark Matter: Relativistc theory for MOND}

\author{\speaker{Jacob D. Bekenstein}\thanks{I dedicate this paper  to Moti Milgrom who started it all.}\\
        Hebrew University of Jerusalem\\
        E-mail: \email{bekenste@vms.huji.ac.il}}

\abstract{MOND, invented by Milgrom, is a phenomenological scheme whose basic premise is that the visible matter distribution in a galaxy or cluster of galaxies alone determines its dynamics.   MOND fits many observations surprisingly well.  Could it be that there is no dark matter in these systems and we witness rather  a violation of Newton's universal gravity law ?  If so, Einstein's general relativity would also be violated.  For long conceptual problems have prevented construction of a consistent relativistic substitute which does not obviously run afoul of the facts.  Here I sketch T$e$V$e\,$S, a tensor-vector-scalar field theory which seems to fit the bill: it has no obvious conceptual problems and has a MOND and Newtonian limits under the proper circumstances.  It also passes the elementary solar system tests of gravity theory.}

\FullConference{28th Johns Hopkins Workshop on Current Problems in Particle Theory\\
		 June 5-8, 2004\\
		 Johns Hopkins University Homewood campus - Bloomberg Center for Physics and Astronomy, Baltimore, Maryland}

\begin{document}

Here I discuss a modified gravity theory. Modified gravity is often mentioned today in relation with dark energy  in cosmology.   But I concentrate on the first systematic modified gravity approach to the mass discrepancy in galaxies and clusters, Milgrom's modified Newtonian dynamics (MOND)~\cite{M1,M2,M3}, and report on a relativistic implementation of it~\cite{B04}.   My approach is phenomenologically motivated, not theory motivated.

\section{\label{sec:intro}Introduction}

There is missing mass in disk galaxies.  We see this from the rotation curve, the plot of the circular orbital velocity with radius.  A few kiloparsecs out from center of a disk galaxy, the rotation curve flattens out.  From the observed stellar distribution Newtonian theory would rather predict a falloff.    Interpreting the rotation curve \emph{a la} Newton tells us that away from the central  regions, the mass enclosed in a given radius  grows proportionately to radius, ultimately giving a galaxy mass an order of magnitude higher than the visible one.  It is as if some invisible matter contributes to the galaxy's gravitation; hence the dark matter (DM) hypothesis.  This finding is accompanied by a mysterious correlation between infrared luminosity of a disk galaxy $L_K$ and the asymptotic  rotational velocity $v_a$: $L_K\propto v_a{}^4$ (Tully-Fisher law)~\cite{SMcG,TF}.   The scatter of the data about the Tully-Fisher law is consistent with observational uncertainties, i.e., there is no intrinsic scatter---a wonderful situation in astrophysics, and one hard to understand from the point of view of the DM paradigm.  

And several lines of evidence show there is missing mass in clusters of galaxies.   Random galaxy motions in a cluster define a (dynamical) velocity scale.   The virial theorem then gives an estimate for the mass interior to the cluster, which comes out much larger than the sum of galaxy masses~\cite{Zwicky,SMcG}.    In the last two decades hot X-ray emitting gas has been found in abundance in many clusters.  From its temperature distribution not only the cluster mass but its distribution can be inferred.  The extra mass in gas has only ameliorated, not removed, the missing mass problem: there is still a factor of $\sim 5$ discrepancy~\cite{SMcG}.  Some clusters are seen to lense distant background galaxies gravitationally.   The measured light deflection has been used to determine the mass and its distribution in each cluster.  Again, the mass comes out larger than the visible mass.  When the dynamical and  lensing methods can be compared,  they give a similar mass discrepancy factor.   In certain circles this is considered a vindication of the DM scenario.

The DM paradigm~\cite{OstrikerPeeblesYahil}, now 30 years old, holds that each galaxy is nested inside a dark halo whose slowly declining mass density profile makes the rotation curve approximately flat over some range.  An halo model necessarily includes at least three free parameters.  They may be the velocity dispersion of the dark particles, the inner cutoff radius (indispensable to prevent mass divergence), and the mass to luminosity ratio  $M/L$ of the visible matter.  But dark clouds hover over this paradigm.  For example, fine tuning of the halo and visible galaxy model parameters is needed to avoid a never-observed cusp in the rotation curve~\cite{BahcallCassertano}.  And attempts to understand the Tully-Fisher law as reflecting galaxy genesis~\cite{Barnes} have problems in explaining how such inherently messy processes can give a sharp correlation~\cite{SReview}.  Add to this the embarrassment of the giant elliptical galaxies; now that extended rotation curves have been determined for some of these, the data reveal no evidence for a dark halo~\cite{Romanowsky}.  And yet it is widely accepted that elliptical galaxies are formed by merger of disk galaxies. One wonders, where the DM is, here or elsewhere ?  No experimental or observational claim of direct discovery of ``dark particles'' has been sustained. 

\section{\label{sec:MOND}The MOND paradigm}

The MOND paradigm is two decades old~\cite{M1,M2,M3}.  Before MOND many workers toyed with the idea that a change in the $1/r^2$ law at large length scales is the explanation for the mass discrepancies~\cite{Morpho}.   But as Milgrom first realized, any modification attached to a length scale would cause the larger galaxies to exhibit the larger discrepancy~\cite{M1,SReview}. This is contrary to the observations: there are small low surface brightness galaxies (LSBs) with large discrepancies, and very large spiral galaxies with small discrepancies~\cite{McGdB98}.   

Milgrom suggested acceleration scale as the more relevant one.  Since then the presence of an acceleration scale in the observations of the discrepancy in spiral galaxies has become more evident~\cite{SReview,McGdB98}.   For example, consider the  log-log plot of the dynamical $M/L_K$ vs. the radius $R$ at the last measured point of the rotation curve for a uniform sample of spiral galaxies in the Ursa Major cluster~\cite{SMcG}.  (Assuming a spherical mass distribution, one calculates $M$ with the Newtonian formula $M=v_a{}^2 R/G$).   The result is a scatter plot.   But if the abscissa  is changed to centripetal acceleration ($a=v_a{}^2/R$ ) at the last measured point, there is a nice correlation:  $M/L_K \propto 1/a$ for $a < 10^{-8} {\rm cm\, s}^{-2}$  and $M/L_K\approx 1$  for $a > 10^{-8} {\rm cm\, s}^{-2}$.  Now  population synthesis models suggest that $M/L_K$ should be about unity~\cite{bell-dejong}.  So it is clear that the Newtonian analysis fails for  $a > 10^{-8}\, {\rm cm\, s}^{-2}$.

MOND maintains that the familiar equality between a test particles's acceleration ${\bf a}$ and the ambient Newtonian gravitational field $-\nabla\Phi_N$  is not exact, and is replaced by
\begin{equation}
\mu(|{\bf a}|/a_0){\bf a}=-\nabla\Phi_N
\label{MOND}
\end{equation}
where $a_0$ is an acceleration scale of order $10^{-8}\,  {\rm cm\, s}^{-2}$, and $\mu(x)\approx x$ for $x\ll 1$ and asymptotes to unity for $x\gg 1$.  So in laboratory settings, where $|{\bf a}|\gg a_0, \ \mu=1$ and we get back to Newtonian physics.  

Now, when $|{\bf a}|\ll a_0$, which occurs frequently in galaxy outskirts, we have the extreme MOND equation $|{\bf a}|{\bf a}/a_0=-\nabla\Phi_N$. For circular orbits on a plane we can set  $|{\bf a}|=v_a{}^2/r$; both sides of the equation then behave  as $1/r^2$ provided $v=$ const. (flat rotation curve).  Equality of the coefficients gives $M=(Ga_0)^{-1}v^4$, or $L_K=(Ga_0\, M/L_K)^{-1} v^4$. On the simple assumption, which accords well with the population synthesis models, that $M/L_K$ is constant, this is precisely the Tully-Fisher law.   Notice that the Tully-Fisher law is derived as a sharp relation, just as observed.

Not only this but MOND has been successful in fitting, for over $10^2$ spiral galaxies of all sizes, the detailed shape of the rotation curves using the observed distribution of stars and gas as inputs~\cite{SMcG}; in these fits the only free parameter is $M/L_K$ (or if more convenient $M/L_B$ in the blue band).   For $90\%$ of the said galaxies there are no significant glitches, and the best fit values for  $M/L_K$ or $M/L_B$ are in excellent agreement with those predicted by the population synthesis models.    Comparable success in fitting rotation curves can be had for dark halo models only by adjusting the above mentioned three halo model free parameters.  MOND is definitely a more parsimonious hypothesis than dark halos. 

We mentioned that the dynamics of some giant elliptical galaxies give no evidence for dark matter.  In MOND this makes sense: these galaxies are found to have accelerations that exceed $a_0$ well out of the core.  To first approximation there should thus not be a deviation of MOND from Newtonian dynamics in them.  Milgrom and Sanders have taken a more precise look at the giant elliptical NGC 3379 (dynamical data from 2003) and shown that it is well described by a MOND based model of 1983 vintage~\cite{MS03}.

But it is not all roses; there are some embarrassments for MOND.  Let us mention three here~\cite{M1,Felten,BekMilg}.  A system of point masses moving \emph{a la} MOND does \emph{not} conserve momentum or angular momentum.   If that were not enough, the motion  of a star orbiting circularly in the outskirts of the the galaxy is predicted to be (extreme) MONDian if the star is regarded as a point mass, but Newtonian if attention is paid to the high accelerations to which the star's gas parcels are subject to.  Finally, open star clusters in the Galaxy, e.g., the Pleiades, are all characterized by internal accelerations small on scale $a_0$, yet display no mass discrepancy whatever, quite at variance with MOND's intent. 

\section{\label{sec:field}Nonrelativistic field theory for MOND}

In light of the above, what are we to make of MOND ?  A sector of the astrophysical community regards it as just a good phenomenological summary of galaxy phenomenology.  But it is clearly more interesting and promising to consider MOND as a reflection of unorthodox physics.  Here we shall consider how far it can be due to modifications of standard gravity theory.  (Alternatively, Milgrom has considered MOND as a modification of Newton's second law~\cite{M1,M94}, but it is unclear to me how such scheme would translate to the relativistic domain).

The first step, a modified nonrelativistic gravity theory, dubbed AQUAdratic Lagrangian theory (AQUAL), solves the mentioned three problems with MOND.  AQUAL is based on the lagrangian~\cite{BekMilg}
\begin{equation}
\mathcal{L} = -{{a}_0{}^2\over 8\pi G}\,
\tilde f\Big({|{\nabla}\Phi|^2\over
{a}_0{}^2}\Big)-\rho\Phi
\label{AQUALlagrangian}
\end{equation}
where $\Phi$ is the true gravitational potential (meaning that for a test particle ${\bf a}=-{\nabla}\Phi$), and $\rho$ denotes the total matter mass density.  The very fact that the formulation is  lagrangian based automatically removes the problems with conservation laws.  

The root of those problems becomes clear when one derives the equation for $\Phi$:
\begin{eqnarray}
{\nabla}\cdot\left[\tilde\mu(|{\nabla}\Phi|/{a}_0){\nabla}\Phi\right]=4\pi G\rho;
\label{AQUAL_Poisson}
 \\
 \quad \tilde\mu(\surd y)\equiv d\tilde f(y)/dy.
\end{eqnarray}
Comparing this with the Poisson equation 
\begin{equation}
\nabla\cdot\nabla\Phi_N=4\pi G\rho
\label{Poisson}
\end{equation}
 we see that $\tilde\mu(|{\nabla}\Phi|/\it a_0)\nabla{\rm \Phi}=\nabla{\rm \Phi}_N-\nabla \times \bf h$, with $\bf h$ a vector field which insures that both sides of the Eq.~Eq.~(\ref{AQUAL_Poisson})  have the same (nonvanishing) curl.  Replacing ${\nabla}\Phi\rightarrow -{\bf a}$ we have the AQUAL equation
\begin{equation}
\tilde\mu(|\bf{a}|/\it a_0)\bf{a}=-\rm \nabla\Phi_N+\nabla \times \bf h.
\label{AQUAL}
\end{equation}
With the proviso that $\tilde\mu(x)$ and $a_0$ here are the same as in Milgrom's scheme, this is a corrected version of the MOND equation~(\ref{MOND}).  The correction $\nabla \times \bf h$ is what brings the dynamics into harmony with the conservation laws, as verified by direct calculation~\cite{BekMilg}.

The same type of calculations show that the center of mass of a collection of mass points moves in the averaged ambient $-\nabla{\rm \Phi}$ in accordance with Eq.~Eq.~(\ref{AQUAL}), regardless of how strong the internal fields acting on the individual mass points are.  This brings agreement with the observation that dense stars and tenuous gas clouds seem to trace out exactly the same rotation curve for each galaxy studied using both.  The AQUAL equation, unlike the original MOND equation~Eq.~(\ref{MOND}), respects the weak equivalence principle (all objects move the same way on a given gravitational field).

The open clusters paradox is also solved by AQUAL in harmony with Milgrom's early conjecture~\cite{M2} that the environment of a system can suppress its MOND-like behavior.  Calculation reveals that the dynamics of a subsystem with internal accelerations below $a_0$ will approach Newtonian dynamics if the enveloping system subjects it to a $-\nabla{\rm \Phi}$ of magnitude exceeding $a_0$~\cite{BekMilg}. This is precisely the situation for the open clusters, many of which are embedded in the gravitational field of the Galaxy where it is not weak on scale $a_0$.  They should thus behave in Newtonian fashion, as found. 

 Further, it is found that if both the internal accelerations and the stronger external field $g_e\equiv |\nabla{\rm \Phi}|$ acting on a subsystem are weak on scale $a_0$, the subsystem's dynamics is again Newtonian, but with an effective gravitational constant $G/\tilde \mu(g_e/a_0)$~\cite{BekMilg}.  Now the dwarf spheroidal companions to the Galaxy, Leo II, Sextans, Draco, etc. have low internal accelerations, and are so far out in the Galaxy's outskirts that they feel a weak external field.  The prediction is then that they should have Newtonian dynamics (which they seem to do in the sense that the dynamically inferred mass distribution mirrors the luminosity distribution), but the $M/L$ determined by virial theorem estimation should be larger than expected from the stellar population involved by a factor $1/\mu(g_e/a_0)$.  And indeed, the dwarf spheroidal exhibit anomalously high virial $M/L$'s, a fact anticipated by Milgrom in his classic papers~\cite{M2}.

For spherical symmetry the vector ${\bf h}$ vanishes; hence for spherical galaxies or clusters the AQUAL formula Eq.~(\ref{AQUAL}) reduces exactly to the MOND formula Eq.~(\ref{MOND}).  We have already mentioned that MOND agrees with the recent finding that the dynamics in giant elliptical galaxies, many of which are quasispherical, do not require dark matter for their explanation when interpreted by Newtonian theory.  We may also recall an early MOND prediction~\cite{M84} that spherical stellar systems with equilibrium isothermal velocity distributions can exist only with typical accelerations $\approx a_0$ or below.  To be sure this is found to be true for the giant molecular clouds in the Galaxy, its dwarf spheroidal satellites and a fair majority of X-ray bright clusters of galaxies.  Giant ellipticals and globular clusters can have higher accelerations, but it is known that the velocity distributions here are anisotropic.  It is a striking finding~\cite{SMcG} that all the five classes of objects have central accelerations within a factor of five of Milgrom's constant $a_0$, originally defined with respect to disk galaxies.    

And what about disk galaxies, MOND's original playground ?  It was known early that the ${\bf h}$ term in Eq.~(\ref{AQUAL}) falls off faster with increasing radius than the $\nabla{\rm \Phi}_N$ term.   Numerical calculations by Milgrom show that in the inner regions ${\bf h}$ contributes only some $10-15\%$ of the acceleration~\cite{Mnumer}.  This means that the old-fashioned MOND reconstruction of rotation curves from the visible matter distribution, which is known to be successful, is very much what a full AQUAL treatment would give.

To be sure MOND and AQUAL have a common failing.  Great clusters of galaxies, which display accelerations up to a few times $a_0$, show an average factor of five discrepancy between the Newtonian dynamical mass and the mass in stars plus gas.  With MOND the discrepancy can go down to a factor of two, but does not go away.  This has to do with the large accelerations in clusters~\cite{SMcG}.  AQUAL does not solve this problem because many of the clusters in question are spherical, making MOND fully equivalent to AQUAL.   And just like MOND, AQUAL does not give a prescription for calculating gravitational lensing by extragalactic objects, so that the mass discrepancies evidenced by the lensing of distant galaxies by clusters cannot be examined critically.  Nevertheless it is clear that in most cases AQUAL theory does well phenomenologically, while having conceptual advantages over the pure MOND prescription.  

\section{\label{sec:steps}Steps towards a relativistic modified gravity}    

Once MOND's philosophy is accepted, a relativistic implementation of it is indispensable, not only for the computation of gravitational lensing, but also to permit investigation of other relativistic systems like close binary pulsars or cosmology.  I now show how hints from AQUAL led to the construction of a good candidate for a relativistic gravitational field theory for MOND~\cite{B04}.

Such theory cannot arise from replacing the Einstein-Hilbert Lagrangian density $R$ of general relativity (GR) by some $ \tilde f(R)$.  True, part of $R$ is quadratic in derivatives of the gravitational potentials (metric components), and such replacement is suggestive of a nonrelativistic limit like Eq.~(\ref{AQUALlagrangian}).   But $R$ also contains a term involving second derivatives of metric components.  In the Einstein-Hilbert action this term reduces to a boundary term and is irrelevant.  But in $\tilde  f(R)$ it would not do so, and would actually dominate the previously mentioned term.   To the above ``poor man's argument'' one can add the Soussa-Woodard theorem~\cite{SW} that a gravitational theory built on metric alone, whether with global or local lagrangian density, cannot both have a MOND like nonrelativistic limit and reproduce the anomalously large gravitational lensing seen in clusters of galaxies.

These problems lend credence to the first relativistic approach to MOND, relativistic AQUAL (RAQUAL) theory~\cite{BekMilg,BCan}.  It promotes $\Phi$ in AQUAL to a scalar field $\phi$, and makes its lagrangian density (henceforth I set $c=1$, but display $G$)
\begin{equation}
\mathcal{L}_{\phi}=-{a_0{}^2\over 8\pi G }
 f\left({g^{\alpha\beta}\phi,_\alpha \phi,_\beta\over a_0{}^2}\right)\end{equation}
where $g^{\alpha\beta}$ is the inverse of the Einstein metric $g_{\alpha\beta}$ and $f$ may be different from $\tilde f$.  One adds to this the  Einstein-Hilbert lagrangian density $R$ as well as a matter lagrangian density ${\cal L}_m$.  This last is obtained from the usual curved spacetime one by the replacement $g_{\alpha\beta}\rightarrow  \tilde g_{\alpha\beta}\equiv e^{2\phi} g_{\alpha\beta} $.  In addition, when integrating ${\cal L}_m$ over spacetime to form the matter action $S_m$, we do it with the determinant of $\tilde g_{\alpha\beta}$.   We call $\tilde g_{\alpha\beta}$ the physical metric.  For example, a point mass $m$ has the action
\begin{equation}
S_m=-m\int e^{\phi}\,(-g_{\alpha\beta}\,
dx^\alpha dx^\beta)^{1/2},
\label{worldlineintegral}
\end{equation}
obtained by replacing $g_{\alpha\beta}\rightarrow  \tilde g_{\alpha\beta}\equiv e^{2\phi} g_{\alpha\beta} $ in the familiar integral over the worldline.  The field equation deriving from $\int \mathcal{L}_\phi(-g)^{1/2}d^4x + S_m$ is [$\mu(y)\equiv f'(y)]$
\begin{equation}
 \left[\mu\left({g^{\alpha\beta}\phi,_\alpha \phi,_\beta\over a_0{}^2}\right)g^{\mu\nu}\phi_{,\nu}\right]=-4\pi G\tilde g^{\alpha\beta}\tilde T_{\alpha\beta}
 \label{RAQUAL}
\end{equation}
where $\tilde T_{\alpha\beta}$ is the energy-momentum tensor obtained by varying the matter action with respect to $\tilde g_{\alpha\beta}$.  For  correspondence with AQUAL we choose $f(y)={\scriptstyle
2\over\scriptstyle 3}y^{3/2}$ for $0<y\ll 1$ and $f(y)\approx
y$ for $y\gg 1$.

How does RAQUAL yield Newtonian and MOND limits where relevant ?  As well known, in the nonrelativistic limit slow motion in GR is governed by the temporal-temporal metric component $g_{tt}$ which can be approximated by $-(1+2{\rm \Phi}_N)$, with ${\rm \Phi}_N$ coming from the usual Poisson equation with $\rho$ as source.  According to Eq.~(\ref{worldlineintegral}) $g_{tt}$'s job is taken over by $\tilde g_{tt}\approx -(1+2{\rm \Phi}_N+2\phi)$ so that it is $\Phi={\rm \Phi}_N+\phi$ which represents the nonrelativistic gravitational potential. Thus $\phi$ will replace the gravitational potential of the dark matter.  For quasistatic situations with nearly flat metric, its field equation (\ref{RAQUAL}) has the form of AQUAL's  Eq.~(\ref{AQUAL_Poisson}).

Now suppose that because of the large scale of the system, $|\nabla\phi|\ll a_0$.  Comparison of $\phi$'s equation with Poisson's Eq.~(\ref{Poisson}) shows, in light of the fact that here $\mu\ll 1$, that $|\nabla\phi|\gg |\nabla{\rm \Phi}_N|$.  Consequently $\nabla{\rm \Phi}\approx \nabla\phi$, so that the physical potential $\Phi$ approximately obeys the AQUAL equation Eq.~(\ref{AQUAL}).  This means that for weak fields (MOND limit), RAQUAL reduces to AQUAL, and inherits the latter's successes.  Now when the system is such that $|\nabla\phi|\gg a_0$ we shall have $\mu\approx 1$ meaning that $\phi$ approximately obeys the equation for ${\rm \Phi}_N$.  Thus $\nabla{\rm \Phi}\approx 2\nabla{\rm \Phi}_N$.  This just mean that at the nonrelativistic level we have Newtonian gravity, but with an effective gravitational constant $G_{\rm eff}=2G$.  We shall not dwell on whether this result is also consistent at the post-Newtonian level because a tenacious problem renders RAQUAL unviable.

Because $  \tilde g_{\alpha\beta}= e^{2\phi} g_{\alpha\beta} $ (conformally related metrics), all optical phenomena, light ray paths included, are exactly alike in the two metrics.  Any hope for anomalously strong light bending, as required by the gravitationally lensed galaxies, hangs on the Einstein metric being substantially different from that in GR with no dark matter.  However, in the case of the clusters $|\nabla\phi|={\cal O}(a_0)$, so that the scalar's energy density is of order $G^{-1}a_0{}^2\sim G^{-1}H_0{}^2$ (here $H_0$ is the present Hubble ``constant'').  This cosmological-scale energy density cannot compete with that furnished by the stellar component in a cluster.  Hence, in RAQUAL $g_{\alpha\beta}$ is not very different from that in GR, and no anomalously strong lensing can be expected from it.  In brief, RAQUAL fails the lensing test miserably.

Sanders and I tried to rescue the RAQUAL idea by replacing the conformal relation between metrics by $\tilde g_{\alpha\beta}=\exp(2\phi)\,[ g_{\alpha\beta} + \varpi\phi,_\alpha\phi,_\beta]$, where $\varpi$ is some function of the invariant $g^{\alpha\beta}\,\phi_{,\alpha}\phi_{,\beta}$~\cite{BS,BKyo}.  Here $\tilde g_{\alpha\beta}$ will bend light rays differently from $g_{\alpha\beta}$.  We only assumed that $\varpi<0$, and this to guarantee that $\tilde g_{\alpha\beta}$ and $g_{\alpha\beta}$ both have the usual signature, and that weak gravitational waves (perturbations of $g_{\alpha\beta}$) propagate inside or on the physical light cone (that of $\tilde g_{\alpha\beta}$).  A calculation then showed that if a cluster can be approximated as static with $\partial\phi/\partial t=0$, then the light bending is \emph{weaker} than in GR for the same mass: the lensing problem becomes worse than in RAQUAL !  

An alternative proposal by Sanders~\cite{Ss} has been more fruitful: in AQUAL multiply just the part of the metric orthogonal to a \emph{fixed} unit time-directed vector ${\frak U}^\alpha$, namely $g_{\alpha\beta} + {\frak U}_\alpha {\frak U}_\beta$, by $e^{-2\phi}$ and that part collinear with ${\frak U}^\alpha$, namely ${\frak U}_\alpha {\frak U}_\beta$, by $e^{2\phi}$, taking the sum of the two parts as the physical metric.  In this so called stratified theory, as in a little known pre-GR paper by Einstein, the ruse immediately produces extra light bending.  Indeed it can resolve the anomalously large lensing problem.  But ${\frak U}_\alpha$  defines a preferred frame, thus breaking local Lorentz invariance, and because ${\frak U}_\alpha$ is specified as time-directed in all situations, the theory cannot be generally covariant.  The first problem is not critical as ${\frak U}_\alpha$ couples to matter only through the physical metric, so that the breakdown of local Lorentz invariance is restricted to the gravitational sector. Sanders found that the breakdown is only weakly reflected in the relevant parametrized post-Newtonian formalism coefficients.  The lack of general covariance is more serious.  This problem cannot be resolved until a prescription for determining ${\frak U}_\alpha$ is given which makes no reference to particular coordinates.  This is where T$e$V$e\,$S takes over.

\section{\label{sec:structure}Structure of the tensor-vector-scalar modified gravity}

This covariant modified gravity theory, dubbed T$e$V$e\,$S for short, starts off from Sanders' prescription for the physical metric:
\begin{equation}
\tilde g_{\alpha\beta}=e^{-2\phi}(g_{\alpha\beta} + {\frak U}_\alpha {\frak U}_\beta) - e^{2\phi} {\frak U}_\alpha {\frak U}_\beta
\label{metric}
\end{equation}
where one demands that $g^{\alpha\beta}{\frak U}_\alpha {\frak U}_\beta=-1$, but does not stipulate any particular orientation for the vector.   The Einstein metric gets its dynamics from the usual Einstein-Hilbert action $S_g$.  The matter action $S_m$ is built by replacing the Minkowsky metric and various derivatives by the physical metric and covariant derivatives with respect to it.  

The vector field's dynamics come from the postulated action
\begin{equation} S_v =-{K\over 32\pi G}\int
\big[g^{\alpha\beta}g^{\mu\nu} 
{\frak U}_{[\alpha,\mu]} {\frak U}_{[\beta,\nu]}
-2(\lambda/K)(g^{\mu\nu}{\frak U}_\mu
{\frak U}_\nu +1)\big](-g)^{1/2} d^4 x;
\label{vector_action}
\end{equation}
the normalization of ${\frak U}_\mu$ is enforced via the Lagrange constraint with multiplier function $\lambda$.  The $K$ is a dimensionless positive coupling constant of the theory.  Note that it is the Einstein metric which appears in $S_v$.  To derive the vector's equation one is directed to vary ${\frak U}_\mu$, as opposed to the contravariant components (${\frak U}^\mu\equiv g^{\mu\nu}U_\nu)$, in $S_v+S_m$.  Now  ${\frak U}_\nu$ enters $S_m$ through the $\tilde g^{\alpha\beta}$, where
\begin{equation}
\tilde g^{\alpha\beta} = e^{2\phi} g^{\alpha\beta} + 2g^{\alpha\mu}g^{\beta\nu}{\frak U}_\mu {\frak U}_\nu \sinh (2\phi).  
\label{inverse}
\end{equation}
This and $(-\tilde g)^{1/2}=e^{-2\phi}(-g)^{1/2}$ have to be varied giving thereby source terms depending on the energy-momentum tensor [$(-\tilde g)^{1/2} \tilde T_{\alpha\beta}=-2\delta S_m/\delta \tilde g^{\alpha\beta}$].  The resulting equation,
\begin{equation}
K{\frak U}^{[\alpha;\beta]}{}_{;\beta}+\lambda
{\frak U}^\alpha+8\pi G\sigma^2
{\frak U}^\beta\phi_{,\beta}g^{\alpha\gamma}\phi_{,\gamma}
= 8\pi G (1-e^{-4\phi}) g^{\alpha\mu} {\frak U}^\beta 
\tilde T_{\mu\beta},
\label{vector}
\end{equation}
is best regarded as giving the value of $\lambda$ together with those of three components of ${\frak U}_\mu$ (the fourth coming from the normalization condition).

Interesting consequences of Eq.~(\ref{vector}) surface if we model matter by a perfect fluid: 
\begin{equation}
\tilde T_{\alpha\beta}=\tilde\rho \tilde u_\alpha\tilde u_\beta
+\tilde p(\tilde g_{\alpha\beta}+\tilde u_\alpha \tilde u_\beta)
\label{oldT}
\end{equation}
Here $\tilde\rho$ is the proper energy density, $\tilde p$ the pressure and $\tilde u_\alpha$ the 4-velocity, all three expressed in the physical metric ($\tilde u^\alpha=\tilde g^{\alpha\beta}\tilde u_\beta$).  It is found that both for a static system, e.g. a quiescent galaxy, and for isotropic cosmology, $U^\mu$ is collinear with the fluid velocity $\tilde u^\alpha$; in particular, they both point in the time direction.  To put it picturesquely, $U^\mu$ ``seeks out'' the frame in which matter is at rest.  In this manner is Sanders' desideratum obtained dynamically rather than by fiat.  

The scalar's action is taken of the form
\begin{eqnarray} S_s &=&-{\scriptstyle 1\over\scriptstyle
2}\int\big[\sigma^2
h^{\alpha\beta}\phi_{,\alpha}\phi_{,\beta}+{\scriptstyle
1\over\scriptstyle 2}G
\ell^{-2}\sigma^4 F(kG\sigma^2) \big](-g)^{1/2} d^4 x,
.\\
h^{\alpha\beta}&\equiv& g^{\alpha\beta}
-{\frak U}^\alpha {\frak U}^\beta
\label{scalar_action}
\end{eqnarray}
where $k$ is another dimensionless positive parameter, $\ell$ is a scale of length, and $F$ is some function. The tensor $h^{\alpha\beta}$ is used instead of $g^{\alpha\beta}$ by itself in order to prevent causality problems (see Sec.~\ref{sec:causality}).  The $\sigma$ in the action is a nondynamical auxiliary field which allows the kinetic term of $\phi$ to be presented as quadratic, which it really is not.  This $\sigma$  is determined by the extremum condition $\delta S_s/\delta\sigma=0$.    The $\phi$ equation is obtained by varying $S_s+S_m$ with respect to $\phi$.  Now $\phi$ enters  $S_m$ through the $\tilde g^{\alpha\beta}$, and variation of this last introduces source terms containing the energy-momentum tensor.  In terms of the function $\mu(y)$ (to be distinguished from Milgrom's $\tilde\mu$) defined by
\begin{equation} 
 -\mu F(\mu) -{\scriptscriptstyle 1\over \scriptscriptstyle 2}\,
\mu ^2F'(\mu ) = y,
\label{F}
\end{equation} 
we obtain the equation
\begin{equation}
\left[\mu\left(k\ell^2 h^{\mu\nu}\phi_{,\mu}\phi_{,\nu}\right)
h^{\alpha\beta}\phi_{,\alpha} \right]_{;\beta}= 
kG\big[g^{\alpha\beta} + (1+e^{-4\phi}) {\frak U}^\alpha
{\frak U}^\beta\big] \tilde T_{\alpha\beta}
\label{s_equation}
\end{equation}
which is plainly reminiscent of the AQUAL equation Eq.~(\ref{AQUAL_Poisson}).

The equations for the metric result from varying $g^{\alpha\beta}$ in $S_g+S_m+S_v+S_s$.  We observe that $S_m$ is built from $\tilde g^{\alpha\beta}$, part of which varies in accordance with Eq.~(\ref{inverse}), and thus introduces the matter's energy-momentum tensor into the equations.   These are 
\begin{equation} 
G_{\alpha\beta}= 8\pi G\Big[\tilde
T_{\alpha\beta} +(1-e^{-4\phi}) {\frak U}^\mu
\tilde T_{\mu(\alpha} {\frak U}_{\beta)}
+\tau_{\alpha\beta}\Big]+
\Theta_{\alpha\beta}
\label{gravitationeq}
\end{equation} where
\begin{eqnarray}
\tau_{\alpha\beta}&\equiv&
\sigma^2\Big[\phi_{,\alpha}\phi_{,\beta}-{\scriptstyle 1\over
\scriptstyle 2}g^{\mu\nu}\phi_{,\mu}\phi_{,\nu}\,g_{\alpha\beta}-
{\frak U}^\mu\phi_{,\mu}\big({\frak U}_{(\alpha}\phi_{,\beta)}-
{\scriptstyle 1\over \scriptstyle
2}{\frak U}^\nu\phi_{,\nu}\,g_{\alpha\beta}\big)\Big]-
{\scriptstyle 1\over\scriptstyle 4}{G\over \ell^2}
\sigma^4 F(kG\sigma^2)  g_{\alpha\beta}
\nonumber
.\\
\Theta_{\alpha\beta}&\equiv&
K\Big(g^{\mu\nu}{\frak U}_{[\mu,\alpha]}
{\frak U}_{[\nu,\beta]} -{\scriptstyle 1\over \scriptstyle 4}
g^{\sigma\tau}g^{\mu\nu}{\frak U}_{[\sigma,\mu]}
{\frak U}_{[\tau,\nu]}\,g_{\alpha\beta}\Big)-
\lambda {\frak U}_\alpha{\frak U}
_\beta
\label{Theta}
\end{eqnarray}

It is most convenient to vary  $g^{\alpha\beta}$ rather than $g_{\alpha\beta}$.  If the latter procedure is implemented cavalierly,   it is easy to get equations which are not the raised indices versions of those coming from the former procedure.  C. Skordis has observed that this happens because in writing Eq.~(\ref{inverse}) as the inverse of Eq.~(\ref{metric}), one is already assuming that $g^{\alpha\beta}{\frak U}_\alpha {\frak U}_\beta=-1$ \emph{before} performing the variation, whereas this normalization should be left free to be enforced by the Lagrange multiplier term.  At any rate, the difference in the equations obtained boils down to a difference in the $\lambda$ appearing in the vector's  $\Theta_{\alpha\beta}$,  but since $\lambda$ has to be computed after the fact from the vector's equation Eq.~(\ref{vector}), and this last is also changed, the difference disappears in the final equations.

It will be observed that $F$ shows up in the scalar's equation and contributes to the source of Einstein's equations.    In my original paper I made the choice
\begin{equation}
F(\mu)={\scriptstyle 3\over \scriptstyle 8}\mu^{-2}\left[\mu \,\left( 4 + 2\,\mu  -
4\,{\mu }^2 + {\mu }^3 \right)  + 2\,\ln [{\left( 1 - \mu  \right) }^2]\right],
\end{equation}
which leads to
\begin{equation} y={\scriptstyle 3\over \scriptstyle 4}\mu^2(\mu-2)^2(1-\mu)^{-1}.
\label{y}
\end{equation}
The important features in this $y(\mu)$ are that for small $y$, $\mu\propto \surd y$ and that large $y$ occurs for $\mu\rightarrow 1$.  There are other choices of $F(\mu)$ that will give these, so the above choice is to be regarded as a toy model.  

Other opportunities for generalizing T$e$V$e\,$S exist.  For example, one can use $\tilde g^{\alpha\beta}$  in place of $g^{\alpha\beta}$ to construct $S_v$ and $S_s$.   Or one can replace the coefficient of ${\frak U}^\alpha {\frak U}^\beta$ in $h^{\alpha\beta}$ by any number below $-1$.  We have not yet elucidated how the theory's predictions would be changed then.   In addition, one can contemplate kinetic terms in $S_v$ containing the symmetric combination ${\frak U}_{(\alpha;\mu)} {\frak U}_{(\beta;\nu)}$ as opposed to  the antisymmetric ones.  But if such a term is added, as in the Jacobson-Mattingly modified gravity with preferred frames~\cite{Jac_Mat}, the gravitational equations then contain second derivatives of ${\frak U}_\alpha$ coming from covariant derivatives; these complicate the formulation of the initial value problem. Finally one could add a kinetic term for $\sigma$, as in the phase coupled gravity (PCG) theory~\cite{BCan,SPCG}.   However, PCG tends to induce instabilities in bound systems~\cite{BRos}, so this addition may be inadvisable.

\section{\label{sec:nonrelativistic}Nonrelativistic limit of T$e$V$e\,$S}

One reason for believing that T$e$V$e\,$S is not a pathological theory is that it has GR as a limit.  Milgrom has remarked that by suitably rescaling  $\phi$ and $\sigma$, one can see that as $\ell\rightarrow\infty$ and $K\rightarrow 0$, the action $S_g+S_m+S_v+S_s$ reduces to its first two terms, with $\lambda$ vanishing and the two metrics becoming identical.  This is pure GR.  Physically this means that for systems small as compared to $\ell$ and for small $K$,  T$e$V$e\,$S should be a passable approximation to standard gravity theory.  We thus suspect that in the real world $K\ll 1$ and the scale $\ell$ is cosmologically large.  A similar comment applies to a second way to get the GR limit (which is established as yet only for static systems and isotropic cosmology):  take $k\rightarrow 0$ with $\ell\propto k^{-3/2}$ and $K\propto k$. 

In the nonrelativistic approximation of T$e$V$e\,$S, one first uses the vector equation Eq.~(\ref{vector}) to compute $\lambda$; one then substitutes it in the linearized Einstein equations, and discards all of $\tau_{\alpha\beta}$ and $\Theta_{\alpha\beta}$ (apart from the $\lambda$ term) because they are quadratic in presumably small quantities like $\phi_{,\alpha}$ or ${\frak U}_{\alpha,\beta}$ (the $F$ term in $\tau_{\alpha\beta}$ also turns out to be negligible after more detailed analysis~\cite{B04}).   In terms of
\begin{equation}
\Xi\equiv e^{-2\phi_c}(1+K/2)^{-1},
\end{equation}
where $\phi_c$ is the asymptotic boundary value of $\phi$ (set by the cosmological model in which the system in question is embedded), one gets by the familiar procedure
\begin{equation}
g_{tt}\approx -(1+2\Xi\Phi_N).
\label{Einstein_metric}
\end{equation}
The coefficient of $\Phi_N$ is not 2 as in GR because $\tilde T_{\alpha\beta}$ appears in the source of Einstein's equations in three distinct forms, only one of which is the familiar one from GR.
After multiplication of $g_{tt}$ by $e^{2\phi}\approx 1+2\phi$,
\begin{equation}
\tilde g_{tt}\approx -(1+2\Phi); \qquad \Phi=\Xi\Phi_N+\phi.
\label{potential}
\end{equation}
The physical gravitational potential $\Phi$ here is quite close to that in AQUAL; recall that we expect $K$ to be small, and it can be shown that $\phi_c$ is small in suitable cosmological models.  Hence $\Xi$ is close to unity.

Let us now compute $\phi$ for a \emph{static} spherically symmetric system with the matter represented by the fluid energy-momentum tensor Eq.~(\ref{oldT}).  Taking $g_{\alpha\beta}$ as flat, the scalar equation Eq.~(\ref{s_equation}) reduces to
\begin{equation}
\nabla\cdot\Big[\mu\Big(k\ell^2 (\nabla\phi)^2\Big)
\nabla\phi\Big]=kG\tilde\rho.
\label{sca_eq}
\end{equation}
Comparing with Poisson's equation Eq.~(\ref{Poisson}) we see that
\begin{equation}
\nabla\phi=(k/4\pi\mu)\nabla  \Phi_N.
\label{int1}
\end{equation}
We may now calculate $\nabla\Phi$ from Eq.~(\ref{potential}) and cast the result as
\begin{equation}
\tilde\mu\nabla\Phi=\nabla\Phi_N; \qquad \tilde\mu=(\Xi+k/4\pi\mu)^{-1}.
\label{int2}
\end{equation}
The first equation is precisely the MOND equation~(\ref{MOND}) since by the nonrelativistic limit of the geodesic equation, and the form of $\tilde g_{tt}$ in Eq.~(\ref{potential}), ${\bf a}=-\nabla\Phi$. Actually in light of Eq.~(\ref{sca_eq})  $\mu$ is regarded as a function of $|\nabla\phi|$.  However, using the last two equations we may reexpress it, as well as $\tilde\mu$, as functions of $|\nabla\Phi|$, just as required by MOND.

Let us now examine the case where because of the weakness of the gravitational field, $\mu\ll 1$.  According to Eq.~(\ref{y}) the auxiliary variable $y$, the argument of $\mu$ in Eq.~(\ref{sca_eq}), is $y\approx 3\mu^2$.  Thus  $\mu\approx (k/3)^{1/2}\ell|\nabla\phi|$.  Using this in Eq.~(\ref{int1}) and Eq.~(\ref{int2}) to eliminate $\nabla\Phi_N$ and $\nabla\phi$ as well as $\tilde\mu$ gives
\begin{equation}
\mu=(k/8\pi\Xi)\big(-1+\sqrt{1+4|\nabla\Phi|/a_0}\,\big);\qquad a_0\equiv (3 k)^{1/2}(4\pi\Xi\ell)^{-1}
\label{plainmu}
\end{equation}
The positive root was chosen to get $\mu>0$.   Of course, this expression for $\mu$ can be believed only when it gives $\mu\ll 1$; assuming $k$ is crudely ${\cal O}(1)$, this will be true when $|\nabla\Phi|\ll a_0$, in which case $\mu\approx (k/4\pi\Xi)|\nabla\Phi|/a_0$. But then Eq.~(\ref{int2}) informs us that $\tilde\mu\approx \Xi^{-1}|\nabla\Phi|/a_0$.  In fact, we get
\begin{equation}
|\nabla\Phi|\nabla\Phi/a_0=\Xi\nabla\Phi_N
\label{final}
\end{equation}
If one identifies $a_0$ in Eq.~(\ref{plainmu}) with Milgrom's constant, this is precisely the extreme ($|{\bf a}|\ll a_0$) limit of the MOND equations~(\ref{MOND}).  The apparently superfluous factor $\Xi$ is understood by referring to Eq.~(\ref{Einstein_metric}) which shows that in T$e$V$e$\,S it  is $\Xi \Phi_N$ which plays the role of nonrelativistic potential in the Einstein metric.  More on this below.

Consider now the case $\mu\approx 1$ which means $y=k\ell^2|\nabla\phi|^2\gg 1$.  It follows from Eq.~(\ref{int2}) that  $\tilde\mu\approx  (\Xi+k/4\pi)^{-1}$.  And that same equation tells us that $\nabla\Phi=(\Xi+k/4\pi)\nabla\Phi_N$.  The proportionality of the two gradients means that the acceleration $-\nabla\Phi$ has Newtonian form, so that $\mu\approx 1$ corresponds to the Newtonian limit of  T$e$V$e$\,S.  And from the coefficient of $\nabla\Phi_N$ we learn that the physical gravitational constant in the Newtonian regime is $G_N=(\Xi+k/4\pi)G$.  The contribution $\Xi G$ has been mentioned already as coming from the gravitational equations; the part $k G/4\pi$ is  contributed by the scalar field.

How big are the departures from exact Newtonian gravity in T$e$V$e$\,S near $\mu=1$ ?  Let us expand $y(\mu)$ about the singular point $\mu=1$ in a Laurent series:
\begin{equation} y={\scriptstyle 3\over \scriptstyle 4} (1-\mu)^{-1} +
\mathcal{O}(1-\mu).
\end{equation}
Because there is no constant term here,  $\mu=1-{\scriptstyle 3\over \scriptstyle 4}y^{-1}=  1-{\scriptstyle 3\over \scriptstyle 4}(k\ell^2|\nabla\phi|^2)^{-1}$ furnishes an accurate inversion.  Eliminating $|\nabla\phi|$ in favor of $|\nabla\Phi|$ with help of Eq.~(\ref{int1}) (where we set $\mu=1$) and  Eq.~(\ref{int2}) gives
\begin{equation}
\tilde\mu\approx {G\over G_N}
\left(1-{16\pi^3\over  k^3} {a_0{}^2\over
|\nabla\Phi|^2}\right).
\label{mutilde}
\end{equation}

In interpreting this result one must realize that if in all formulae involving observables, the constant $G$ is reexpressed in terms of $G_N$, the prefactor $G/G_N$ here will disappear.  Thus the leading term in $\tilde\mu$  is really unity, as in MOND.  Next we have a correction which still relates to nonrelativistic gravity, and vanishes as $|\nabla\Phi|/a_0\rightarrow\infty$.  In a strong gravity system, i.e., the solar system, this correction is small enough not to bring about conflict with the very accurate ephemeresis if  $k={\cal O}(1)$.  But a much smaller $k$ may cause trouble.  So far the only reason for being interested in small $k$ is cosmology, where small $k$ leads to easy to understand models.  But if good models cannot be had for $k={\cal O}(1)$, one would have to rethink the form of $F(\mu)$ and perhaps other features of the theory.

The cases $\mu\ll 1$ and $\mu\approx 1$ conform with the pure MOND equation~(\ref{MOND}).  But when $\mu$ has an intermediate value, T$e$V$e$\,S introduces corrections to MOND.  In this regime $\Phi_N$ and $\phi$ make comparable contributions to $\Phi$.  But they satisfy two different equations,  (\ref{AQUAL_Poisson}) and (\ref{Poisson}), which means that in general $\Phi$ is not exactly a solution of a MOND type equation.

As I mentioned in Sec.~\ref{sec:field}, there are clusters of galaxies whose dynamics are not well described by MOND.   One possible reason may be that such clusters harbor still unseen matter~\cite{S99}.  But the above mentioned correction offers an alternative.   Even for spherical clusters T$e$V$e$\,S predicts deviations from pure MOND behavior, particularly in those systems with typical internal accelerations near $a_0$.  This possibility has not yet been explored quantitatively. 

\section{\label{sec:relativistic}Relativistic corrections in T$e$V$e$\,S} 

The post-Newtonian effects have often been used to rule out competitors of GR; how does T$e$V$e$\,S fare in this respect ?   Here I briefly sketch how to calculate from T$e$V$e$\,S the parametrized post-Newtonian coefficients $\beta$ and $\gamma$ for the physical metric exterior to the Sun,
\begin{eqnarray} \tilde g_{\alpha\beta} dx^\alpha dx^\beta&=&-e^{\tilde \nu}
dt^2+e^{\tilde\varsigma} [d\varrho^2+
\varrho^2(d\theta^2+\sin^2\theta d\varphi^2)],
\label{spherical}
\\
e^{\tilde \nu}&=&1-2G_N\, m\, \varrho^{-1}+2\beta G_N{}^2 m^2
\varrho^{-2}+\mathcal{O}({\varrho^{-3}}),
\\
e^{\tilde\varsigma}&=&1+2\gamma G_N\, m\,
\varrho^{-1}+\mathcal{O}({\varrho^{-2}}).
\end{eqnarray}
Here $m$ is the Sun's physical mass and $G_N$ was defined previously.   The $\gamma$ determines the light deflection and radar signals time delay, while both it and $\beta$ determine the perihelion precessions of the various planets.

The first stage is to write the Einstein metric in analogy with Eq.~(\ref{spherical}) with  the metric functions expanded as
\begin{eqnarray} e^\nu&=&1-r_g/\varrho+\alpha_2 (r_g/\varrho)^2
+\cdots
\label{expansion1}
.\\ e^\varsigma&=&1+\beta_1 r_g/\varrho+\beta_2 (r_g/\varrho)^2
+\cdots\
\label{varsigma}
\end{eqnarray}
The $r_g$ is a scale of length to be determined, and so one coefficient can be taken as $-1$ (negative because gravity is attractive).  Symmetry and normalization dictate that 
\begin{equation}
{\frak U}^\alpha=\{e^{\nu/2},0,0,0 \}.
\label{vec}
\end{equation}
With this one solves the scalar equation to get
\begin{equation}
\phi(\varrho)=\phi_c-{kGm_s\over 4\pi \varrho}+\mathcal{O}({\varrho^{-3}}).
\label{intscalar}
\end{equation}
Again, $\phi_c$ is the asymptotic value of $\phi$ and $m_s$ is a particular integral over $\tilde\rho$ and $\tilde p$ of the sun's matter which is close to the everyday formula for mass~\cite{B04}.

Next one computes outside the Sun the parts of $\tau_{\alpha\beta}$ and $\Theta_{\alpha\beta}$  which fall off as $\varrho^{-4}$.  Using them one solves the $G_t{}^t$ and  $G_\varrho{}^\varrho$ components of the gravitational equations Eq.~(\ref{gravitationeq}) by power series to determine the coefficients (with errors in Ref.~\cite{B04} corrected) 
\begin{equation}
\beta_1=1;\qquad 
\alpha_2={\scriptstyle 1\over \scriptstyle 2};\qquad
\beta_2={\scriptstyle 3\over \scriptstyle 8}+{\scriptstyle 1\over
\scriptstyle 16}K-{kG^2m_s{}^2\over 8\pi r_g{}^2}.
\label{beta_2}
\end{equation}
The next stage is to match the interior and exterior solutions for $e^\varsigma$ at the Sun's surface, which procedure determines that $r_g=2Gm_g+{\cal O}(r_g{}^2/R)$  where $R$ is the Sun's radius and $m_g$ is an integral over the matter variables, $\tau_{tt}$ and $\Theta_{tt}$, which agrees with the na\" ive  expression for mass in the Sun (as does $m_s$) to  $10^{-5}$ fractional precision. 

Finally one uses Eqs.~(\ref{metric}), (\ref{vec}) and (\ref{intscalar}) to calculate the physical metric.  The $e^{\pm 2\phi_c}$ factors entering in transformation Eq.~(\ref{metric}) are gotten rid of by redefining units of length \emph{and} time.  The result can be put in form  Eq.~(\ref{spherical}) with the values (corrected for an error in Ref.~\cite{B04})
\begin{equation}
\gamma=1;\qquad \beta=1 ,
\end{equation}
where $m=(G/G_N)(m_g+km_s/4\pi)$ is the  observable mass.  Both $\beta$ and $\gamma$ are here the same as in GR.  The classical solar systems tests cannot thus distinguish between the theories with current measurements precision, and
T$e$V$e$\,S passes the elementary post-Newtonian tests.  A desirable thing for the future is a calculation for  T$e$V$e$\,S of the three post-Newtonian coefficients which have to do with preferred frame effects, and are today well constrained by experiment.

\section{\label{sec:lensing}Gravitational lensing according to  T$e$V$e$\,S}

We recall that lensing by clusters of galaxies gives evidence that the same mass is responsible for galaxy dynamics in and gravitational lensing by the clusters.  If there is DM with suitable distribution, this accords well with GR.  As we now show, T$e$V$e$\,S does just as well~\cite{B04}.

It is sufficient to work in linearized gravity.   Write the  metric as 
\begin{equation}
g_{\alpha\beta}=\eta_{\alpha\beta}+\bar
h_{\alpha\beta}-{\scriptscriptstyle 1\over \scriptscriptstyle 2}\,
\eta_{\alpha\beta} \,
\eta^{\gamma\delta} \bar h_{\gamma\delta}
\end{equation}
where the perturbation from flat spacetime, $\bar h_{\gamma\delta}$, satisfies the Lorentz gauge condition
\begin{equation}
\eta^{\beta\delta}\partial _\beta\bar h_{\gamma\delta}{} =0.
\end{equation}
Then the Einstein tensor to ${\cal O}(h)$ takes the form
\begin{equation}
G_{\alpha\beta}=-{\scriptscriptstyle 1\over
\scriptscriptstyle 2}\,
\eta^{\gamma\delta}\partial_\gamma\partial_\delta
\,\bar h_{\alpha\beta}.
\end{equation}
Accordingly, the gravitational equations~(\ref{gravitationeq}) take the form
\begin{equation}
\eta^{\gamma\delta}\partial_\gamma\partial_\delta
\,\bar h_{\alpha\beta}= -16\pi G\Big[\tilde
T_{\alpha\beta} +(1-e^{-4\phi}) {\frak U}^\mu
\tilde T_{\mu(\alpha} {\frak U}_{\beta)}
+\tau_{\alpha\beta}\Big]-2
\Theta_{\alpha\beta}
\label{grav}
\end{equation}

Galaxies and clusters of galaxies are quiescent systems, so we may drop time derivatives of $\bar h_{\alpha\beta}$; likewise, we drop the quadratic tensors $\tau_{\alpha\beta}$ and $\Theta_{\alpha\beta}$ apart from the term $-\lambda{\frak U}_\alpha {\frak U}_\beta$ which is not ostensibly small.  The $\lambda$ is calculated from Eq.~(\ref{vector}) after neglecting the $\phi$ terms.  For $\tilde
T_{\alpha\beta}$ we take the form Eq.~(\ref{oldT}).  Not surprisingly we recover an expression for $\bar h_{tt}$ we already met in Sec.~\ref{sec:nonrelativistic}: $\bar h_{tt}=-4\Xi\Phi_N$, where $\Phi_N$ comes from Poisson's equation with $\tilde\rho$ as  source, and $\Xi$ has been defined earlier.  Now in nonrelativistic systems $\tilde p\ll \tilde\rho$ while velocities are small, so that the spatial-spatial and spatial-temporal components of the source of Eq.~(\ref{grav}) are small compared to that of the temporal-temporal one.  We can thus claim that $\bar h_{ij}\approx 0$ and $\bar h_{ti}\approx 0$, so that 
\begin{equation}
g_{\alpha\beta}=(1-2\Xi\Phi_N)\eta_{\alpha\beta}-4\Xi\Phi_N\delta_\alpha^t\delta_\beta^t.
\end{equation}
We now use Eq.~(\ref{metric}) to linear order in $\phi$ to construct the physical metric.
The corresponding line element is
\begin{equation} \tilde g_{\alpha\beta} dx^\alpha dx^\beta
=-(1+2\Phi)dt^2+(1-2\Phi)\delta_{ij} dx^i dx^j
\label{weakmetric}
\end{equation}
where again $\Phi=\Xi\Phi_N+\phi$.

Metric Eq.~(\ref{weakmetric}) is the usual linearized metric used in GR to calculate extragalactic lensing. So just as in GR, in T$e$V$e\,$S a single potential is responsible both for galactic dynamics and for lensing.   This solves the principal observational problem of early modified gravity theories in regard to lensing (Sec.~\ref{sec:steps}).  Of course within GR  $\Phi$ comes from the Poisson's equation with DM included in the source, whereas in T$e$V$e$\,S  both the $\Phi_N$ and $\phi$ parts of $\Phi$ have only visible matter as sources, and the latter is computed from Eq.~(\ref{AQUAL_Poisson}) which is a suitable limit of Eq.~(\ref{s_equation}).   These two prescriptions for $\Phi$ need not agree \emph{a priori}, but as we have argued,  nonrelativistic low acceleration dynamics in T$e$V$e\,$S are approximately of MOND form, and MOND's predictions have been found to agree with much of galaxy dynamics phenomenology.  We thus expect T$e$V$e\,$S's predictions for gravitational lensing by galaxies (and some clusters of galaxies) to be in no way inferior to those of GR dark halo models.    Now the MOND formula Eq.~(\ref{MOND}), and T$e$V$e\,$S, at least with our choice of $F$,  both predict that asymptotically the potential $\Phi$ of an isolated  galaxy grows logarithmically with distance \emph{indefinitely} provided the environmental effect discussed in Sec.~\ref{sec:field} is not important.  Dark halo models do not.  So T$e$V$e\,$S for a specific choice of $F$ is in principle falsifiable.  Dark matter is less falsifiable because of the essentially unlimited choice of halo models and choices of their free parameters.  

\section{\label{sec:causality}The problem of causality}

Not only did the RAQUAL theory of MOND fail on account of its lack of specific gravitational lensing, but it also exhibited superluminal scalar waves.  In fact, modified gravity of any sort with a scalar sector is easily afflicted by such acausality~\cite{BRos}.  T$e$V$e\,$S, although designed for other purposes, resolves this problem neatly.  

To show the root of the problem and its solution, I will focus on the scalar equation, Eq.~(\ref{s_equation}), in a region outside the sources.  Separating $\phi$ background and perturbation as $\phi=\phi_{\rm B}+\delta\phi$, and linearizing in $\delta\phi$ gives
\begin{equation}
 \left(g^{\alpha\beta}-{\frak U}^\alpha {\frak U}^\beta+2\xi H^\alpha
H^\beta\right)\partial_\alpha\partial_\beta\, \delta\phi + \cdots=0,
 \label{linear}
\end{equation}
where $H^\alpha\equiv (h^{\mu\nu}\partial_{,\mu} \partial_{,\nu} \phi)^{-1/2}h^{\alpha\beta}\partial_{,\beta}\phi$, $\xi\equiv d\ln \mu(y)/d\ln y$, the $\cdots$ stand for terms with one $\phi$ derivative (for which reason we display only plain and not covariant derivatives), and the subscript $B$ is dropped here and henceforth.  If we work with RAQUAL's Eq.~(\ref{RAQUAL}), the only difference is that the term $-{\frak U}^\alpha {\frak U}^\beta$ is missing in Eq.~(\ref{linear}) and in the definition of $H^\alpha$.  In a static situation $\partial_t\phi=0$, and as mentioned in Sec.~\ref{sec:structure}, ${\frak U}^\alpha$ has only a time component; hence for both theories we may replace $h^{\alpha\beta}\rightarrow g^{\alpha\beta} $ in  $H^\alpha$, which means this vector is purely spatial with unit norm with respect to $g_{\alpha\beta}$.

Let us look at \emph{short wavelength} $\phi$ perturbations of RAQUAL in a local Lorentz frame of the \emph{physical} metric $\tilde g_{\alpha\beta}$; the frame is  oriented so that
\begin{equation}
H^\alpha=e^{\phi}\{0,1,0,0\},
\end{equation} 
where the $e^\phi$ is needed because $\tilde g_{\alpha\beta} H^\alpha H^\beta = e^{2\phi} g_{\alpha\beta} H^\alpha H^\beta = e^{2\phi}$.  For $g^{\alpha\beta}$ we must evidently write in Eq.~(\ref{linear}) $e^{2\phi}\eta^{\alpha\beta}$.   Making a WKB approximation, which amounts to replacing $\partial_\alpha\partial_\beta\, \delta\phi\rightarrow -k_\alpha k_\beta \, \delta\phi$ ($k_\alpha$ is the wavevector), and discarding  the much smaller $\cdots$ term, we get the dispersion relation
\begin{equation}
\omega = -k_t =[(1+2\xi)k_x{}^2+k_y{}^2+k_z{}^2]^{1/2}.
\end{equation}
The group speed $v_{\rm g}=|\partial
\omega/\partial
\mathbf{k}|^{1/2}$ turns out to be anisotropic:
\begin{equation} v_{\rm g}=\left[{(1+2\xi)^2 k_x{}^2+k_y{}^2+k_z{}^2
\over (1+2\xi) k_x{}^2+k_y{}^2+k_z{}^2}\right]^{1/2}.
\end{equation}   

In the deep MOND regime $[f(y)={\scriptstyle
2\over\scriptstyle 3}y^{3/2}$],
$2\xi=1$ while in the high acceleration limit [$f(y)\approx
y$], $\xi\approx 0$.   Thus whatever the choice of $f$,  $0<\xi<1$ over some range of $y$ (acceleration).  There $v_{\rm g}>1$ unless $k_x=0$.  This is no artifact of the units chosen; the conformal relation between $\tilde g_{\alpha\beta}$ and $g_{\alpha\beta}$ preserves ratios of space and time coordinates, and hence leaves velocities invariant.  Now light waves travel on light cones of $\tilde g_{\alpha\beta}$ while metric waves do so on null cones of $g_{\alpha\beta}$.  Since two metrics are conformally related, so their null cones coincide, light and metric waves travel with unit speed.  Thus most $\phi$ waves are superluminal, in violation of the causality
principle.  It may be seen that the blame rests on the vector $H^\alpha$ which, of course, cannot be avoided within this structure of RAQUAL.
 
Now turning to T$e$V$e$\,S we rewrite Eq.~(\ref{linear}) with help of Eq.~(\ref{inverse}) as
\begin{equation} [e^{-2\phi}\tilde g^{\alpha\beta}-(2-e^{-4\phi}){\frak U}^\alpha {\frak U}^\beta+2\xi
H^\alpha H^\beta]\,\partial_\alpha\partial_\beta\, \delta\phi+ \cdots=0.
\label{modwaveeq}
\end{equation}
In the local Lorentz frame of $\tilde g_{\alpha\beta}$ whose time and $x$ axes are oriented with ${\frak U}^\alpha$ and $H^\alpha$, respectively, we must put $\tilde g^{\alpha\beta}\rightarrow \eta^{\alpha\beta}$.  Normalizing ${\frak U}^\alpha$ and $H^\alpha$ with respect to $g_{\alpha\beta}$ means taking
\begin{equation}
H^\alpha=e^{-\phi}\{0,1,0,0\};\qquad {\frak U}^\alpha=e^{\phi}\{0,1,0,0\}.
\end{equation}
Again making a WKB approximation gives the new dispersion relation
\begin{equation}
\omega = -k_t =(2)^{-1/2}e^{-2\phi}[(1+2\xi)k_x{}^2+k_y{}^2+k_z{}^2]^{1/2}.
\end{equation}
From this we get the group velocity
\begin{equation}
v_{\rm g}={e^{-2\phi}\over \surd 2}\left[{(1+2\xi)^2
{\kappa}_x{}^2+{\kappa}_y{}^2+{\kappa}_z{}^2
\over(1+2\xi)
{\kappa}_x{}^2+{\kappa}_y{}^2+{\kappa}_z{}^2}\right].
\end{equation}

The $\xi$ parameter is computed from Eq.~(\ref{y}):
\begin{equation}
\xi(\mu)=(\mu-1)(\mu-2)/(3\mu^2-6\mu+4)
\end{equation}
from which follows that in the range $0<\mu<1$ corresponding to $y>0$ (recall, in the static case $y=k\ell^2|\nabla\phi|^2$), $0<\xi<{\scriptstyle 1\over \scriptstyle 2}$ with the lower bound corresponding to the Newtonian regime while the upper bound is approached in the extreme MOND regime.  Consequently, in the deep
MOND regime,  $v_{\rm g}\leq e^{-2\phi}$ with equality for
$k_y=k_z=0$.   In the Newtonian regime  $v_{\rm g}=2^{-1/2}
e^{-2\phi}$ for all ${\bf k}$.  Finally, in the intermediate
regime  $2^{-1/2} e^{-2\phi}\leq v_{\rm g}\leq  (1+2\xi)^{1/2}2^{-1/2}
e^{-2\phi}$, with lower and upper equality for
$k_x=0$ and $k_y=k_z=0$, respectively.  Since cosmological models with $\phi$ positive and small throughout can be constructed in T$e$V$e$\,S, and static systems embedded in them have a $\phi$ with the same characteristics~\cite{B04}, $\phi$ waves can propagate causally.

Causal propagation is maintained in a cosmological background.  In addition, it is maintained for metric and vector perturbations both on static backgrounds and in cosmological ones~\cite{B04}.  Thus in contrast with RAQUAL, T$e$V$e$\,S is consistent with causality.

\section{Conclusions}

Many a lecture on galaxy phenomenology pointed out that there does not exist a relativistic formulation of MOND, and that consequently MOND does not have to be taken seriously as new physics.   More than anything, the present talk makes it clear that there is nothing that prevents a covariant formulation of MOND at the phenomenological level.  T$e$V$e$\,S not only reproduces the MOND paradigm in the appropriate regime.   It also passes the elementary post-Newtonian solar system tests, and removes a glaring observational problem with gravitational lensing by extragalactic systems.  And it resolves the problems of superluminal propagation that accrued to aspirants to the title of relativistic MOND.  But some problems, such as the failure to achieve a perfect Newtonian limit in the outer solar system exist.  There remains a large labor to assess how these may be fixed, and to extract consequences of T$e$V$e$\,S for the study of cosmological perturbations, gravitational wave astronomy, binary pulsar timing, and post-Newtonian tests regarding preferred frame effects, to name the most obvious.

\acknowledgments

Many hours of discussion with Moti Milgrom and Bob Sanders have shaped my understanding of MOND.  Thanks also to Constantinos Skordis for insights on deriving the gravitational equations in constrained systems, and to Dimitrios Giannios for catching an algebraic error in the calculations of Ref.~\cite{B04} for  the PPN coefficients.  This research was supported by the Israel Science Foundation (grant No. 694/04).

\end{document}